\documentclass[a4paper]{article}

\usepackage{INTERSPEECH2020}
\usepackage{makecell}

\title{Improving on-device speaker verification using federated learning with privacy}
\name{Filip Granqvist, Matt Seigel, Rogier van Dalen, \'{A}ine Cahill, Stephen Shum, Matthias Paulik}
\address{Apple}
\email{\{fgranqvist,mseigel,rogier\_vandalen,aine\_cahill,stephen\_shum,mpaulik\}@apple.com}

\begin{document}

\maketitle
\begin{abstract}
Information on speaker characteristics can be useful as side information in improving speaker recognition accuracy. However, such information is often private.
This paper investigates how privacy-preserving learning can improve a speaker verification system, by enabling the use of privacy-sensitive speaker data to train an auxiliary classification model that predicts vocal characteristics of speakers.
In particular, this paper explores the utility achieved by approaches which combine different federated learning and differential privacy mechanisms.
These approaches make it possible to train a central model while protecting user privacy, with users' data remaining on their devices.
Furthermore, they make learning on a large population of speakers possible, ensuring good coverage of speaker characteristics when training a model.
The auxiliary model described here uses features extracted from phrases which trigger a speaker verification system. From these features, the model predicts speaker characteristic labels considered useful as side information.
The knowledge of the auxiliary model is distilled into a speaker verification system using multi-task learning, with the side information labels predicted by this auxiliary model being the additional task.
This approach results in a 6\,\% relative improvement in equal error rate over a baseline system.
\end{abstract}

\section{Introduction}\label{sec:intro}

Speaker verification is the problem of determining whether the person speaking is a specific individual or someone else.
It is a vital feature for devices that use a ``wake-up phrase'' to provide access to information, as actions should only be triggered when this phrase is uttered by the device owner and not an impostor.
Speaker verification systems usually consist of two components: a speaker embedding network; and a discriminative method for comparing pairs of embeddings to determine whether or not those embeddings originate from the same speaker \cite{variani14,snyder17,zhu18,snyder18}.
Additional side information can be useful for speaker verification \cite{ferrer08,plchot13,kelly15}.
This side information could be obtained through manual labelling.
The setting that this paper considers instead is one where side information is available on many users' devices, but it is privacy-sensitive and should therefore not be uploaded to a central server.
At a high level, this paper tests three hypotheses:
\begin{enumerate}
  \item It is possible to train a classifier on the audio of trigger phrases to predict personal attributes of the speaker considered to be useful as side information.
  \item Such a classifier can be improved with federated learning while preserving users' privacy.
  \item The predictions of this classifier can be used to improve the performance of speaker verification.
\end{enumerate}

Previous work has shown that neural networks can learn to predict speaker-dependent labels, such as gender \cite{wang17,meinedo10,kabil18}, and emotion \cite{han14,ghosh16,wang17}, from utterances.
The desired outcome from testing the first hypothesis is a classifier that can predict similar speaker-dependent labels from the same input as the baseline speaker verification system used in this paper.

The second hypothesis is that it is possible to train a useful classifier on distributed user data while preserving user privacy. This is achieved through the combination of federated learning with differential privacy, which has been proposed and put into practice successfully in a large body of prior work \cite{abadi16,mcmahan16,mcmahan18,ryffel18}.  
In federated learning, a batch of clients compute statistics on their local data using the latest version of a central model. The resulting statistics are combined on a server to improve the central model. This process is repeated with a different subset of users. Federated averaging \cite{mcmahan16} is commonly used for federated learning. In this algorithm, models are trained locally on devices and the changes in model parameter values are averaged on a central server and used to update the central model.
However, local model updates, which are derived from the data, might leak sensitive information. To prevent this, differential privacy (DP) \cite{dwork14} is used in this paper.
Prior work has provided few examples of high-utility applications on real-world models and datasets, and none on classifying speakers. This paper presents an analysis of different privacy regimes on training accuracy and convergence in this domain.

The third hypothesis states that the encoded knowledge of an auxiliary model trained on side information can be used to improve a speaker verification system.
Manually labelled side information has been shown effective for improving speaker verification systems \cite{ferrer08,plchot13,kelly15}.
The baseline system \cite{phs18} employs the common approach of using a speaker embedding network and scoring pairs of embeddings using cosine similarity.
This paper shows that it can be improved by enriching the speaker embedding network with knowledge distilled from the auxiliary model.

The structure of this paper is as follows.
Section~\ref{sec:pfl} provides a high-level overview of how federated learning can be made private using differential privacy.
Section~\ref{sec:phs} introduces the baseline speaker verification system used in production.
Section~\ref{sec:vc} introduces the classifier trained on user data in a privacy-preserving manner to predict side information from trigger phrases, named the ``vocal classification model''.
Section~\ref{sec:setup} explains the approach for including the vocal classification model in the speaker verification training setup to ultimately improve performance.
Section~\ref{sec:results} presents experimental results.

\section{Federated learning with privacy}\label{sec:pfl}

Data that can be used to improve machine-learned models often belongs to individuals or users and is therefore distributed over their devices. Federated learning is an approach that makes learning in this scenario possible. In federated learning, a central model is trained over a distributed dataset, where a large number of nodes (e.g. user devices) hold variable-sized subsets of the data. A model update or gradient  \cite{duchi12, konecny16, mcmahan16} is computed at the node on the local data, and communicated to a central server. A large number of these updates or gradients are combined at the central server during each iteration of training.
A global update to the central model is computed as the average of local updates.
This is called ``federated averaging'' \cite{mcmahan16}.

Many organisations and policy makers are committed to upholding user privacy. This makes federated learning an important approach to consider when dealing with data that is private, as it goes some way to protecting privacy.
However, even though raw user data is not communicated with the server, it has been shown that model updates can leak information about the raw data \cite{fredrikson15, melis18}. As mentioned in Section \ref{sec:intro}, there is a large body of work investigating and putting approaches into practice which combine federated learning with some privacy protection. One common way to mitigate these threats to privacy is to apply differential privacy (DP) \cite{dwork14,abadi16,mcmahan18,ryffel18}. Differential privacy makes it possible to add noise to the model updates to give a guaranteed upper bound on the amount of information that can be leaked.
DP can be used to protect an individual's update by applying noise at the distributed node. 
DP can also be applied centrally to protect the privacy of individuals' updates after aggregation \cite{shokri17}.

In this paper, a number of privacy regimes are explored in simulation. One of these regimes is to use a weaker form of local DP \cite{bhowmick19}, combined with central DP. This form of local DP is applied to individual updates which are sent for aggregation on a secure server. This algorithm is an optimal method for providing updates (i.e. high-dimensional vectors) with the highest possible signal-to-noise ratio (SNR). The algorithm is tuned to achieve an SNR that permits high accuracy, while still allowing strong DP guarantees in deployment scenarios where it is applicable, e.g., because of shuffling \cite{erlingsson19} and subsampling \cite{balle18subsampling}.
In addition, while doing federated averaging, the server adds enough additional noise to ensure strong central DP guarantees (as per the moment's accountant \cite{abadi16}.

\section{Speaker verification}\label{sec:phs}
The baseline speaker verification system improved in this work is the system described in \cite{phs18}, which builds on an underlying voice trigger model that recognizes the trigger phrase. The input of the speaker verification system is a fixed-length supervector with features generated from forward propagation of the trigger phrase audio through the voice trigger model.
The voice trigger system uses 26 Mel Frequency Cepstral Coefficients to parameterize a hidden Markov model (HMM) which models the trigger phrase.
The means of the 31 HMM states (other than those modelling silence) are concatenated to form the supervector, resulting in $26\times20=520$ dimensions.

The supervector consists of features about the particular trigger phrase and the baseline speaker verification system contains a neural network that transforms the supervector into ``speaker space'', only focusing on retaining characteristics about the speaker itself. 
The neural network is a fully connected neural network with five layers. 
The first four layers consist of $256$ dimensions, with batch normalization \cite{ioffe15} and sigmoid activations. 
The fifth layer is designed to be the embedding layer, and is therefore only a linear transform of dimension 100 with batch normalization. 
In the training phase, a sixth layer of size $K$ with softmax activation is added as a head to the architecture, where $K$ is the number of speakers in the training dataset. 
Training is defined as a speaker identification task, where the labels are one-hot representations of the $K$ unique speakers and training is performed by minimizing the cross-entropy with the output softmax distribution.
The trained embedding is used for measuring the similarity between two utterances by measuring the distance in ``speaker space''.

The speaker verification system stores multiple speaker vectors generated from a set of enrolment utterances. At test time, the acoustic instance of a trigger phrase is transformed into a fixed-length supervector from the HMM states of the voice trigger model, transformed again into a speaker vector by the speaker embedding network, and compared to the enrolment speaker vectors using cosine similarity.
The average cosine similarity between the test speaker vector and the enrolment vectors can be interpreted as a speaker verification score.
Following the notation of \cite{marchi18}, the score is defined as

\vspace*{-0.2cm}
\begin{equation}
  SV_{score} = \frac{1}{N}\sum_{i=1}^N  \frac{f_{nn}(u_a)^\top f_{nn}(u_i^{spk}) }{\Vert f_{nn}(u_a) \Vert \Vert  f_{nn}(u_i^{spk}) \Vert }
  \label{eq:predict}
\end{equation}
where $u_i^{spk}$ is the $i$th out of $N$ supervectors from the enrolment phase, $u_a$ is the test supervector, and $f_{nn}$ is the function expressed by the speaker embedding network. 
The speaker verification score $SV_{score}$ is then compared to the operating threshold $\lambda$ to reject or accept the request following the trigger phrase. 

The work presented in this paper mainly targets the Japanese language (\texttt{ja\_JP}), where the training dataset originates from a speaker population of size $K=18700$. The speaker embedding network is trained with a batch size of $256$, a weight decay of $5\cdot10^{-4}$, initial learning rate of $10^{-4}$ and momentum of factor $0.9$. The performance of this setup is presented as \textit{Baseline} in Section \ref{sec:results}. Improvements on the baseline model presented in the paper are also trained with the same hyperparameters unless otherwise stated.

\section{Vocal classification}\label{sec:vc}
The first hypothesis proposes that classification of side information can be learned from the voice trigger phrase only, if the side information is correlated with vocal characteristics.
To test the hypothesis, a fully connected deep neural network was defined which uses the same input features as the baseline speaker verification system, i.e.\ the 520-dimensional supervector, and predicts the side information of the speaker. 
This model is referred to as the ``vocal classification model''.
The target labels are sensitive information and are stored privately on devices, hence larger-scale experiments of training the vocal classification model were only possible using the framework explained in Section \ref{sec:pfl}.  

\begin{figure}[!htb]
\centering
\includegraphics[scale=0.09]{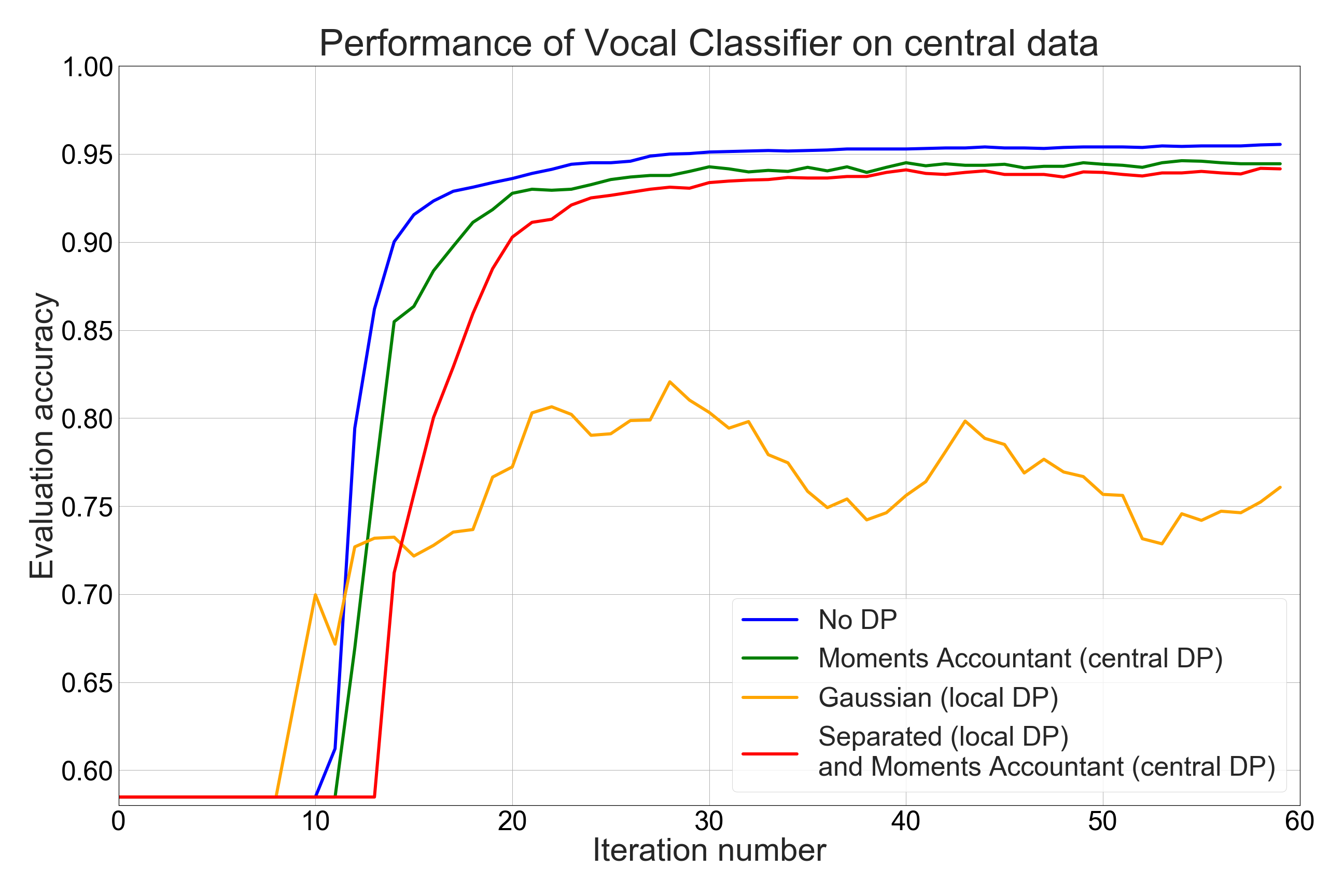}
\vspace*{-0.7cm}
\caption{Accuracy of the vocal classification model on an evaluation dataset, trained with different DP mechanisms.}
\label{fig:vc}
\end{figure}

Experiments were carried out with limited central data to evaluate the effect of applying differential privacy on federated training of the vocal classification model.
In addition to accuracy, the signal-to-noise ratio is measured throughout experiments to quantify the signal quality of model updates where DP has been applied. Here, SNR is computed as the ratio of the L2 norm of the un-noised model update to that of the added DP noise, i.e. $\mathrm{SNR} = \frac{||\mathrm{unnoised~update}||_2}{|| \mathrm{DP~noise} ||_2}$.
Figure \ref{fig:vc} shows the accuracy of the following experiments in simulation on an evaluation set:

\begin{description}
  \item[No DP] No DP mechanism is applied. The resulting accuracy of $95.6\%$ acts as an upper bound for the remaining experiments because introducing any DP mechanism to improve privacy guarantees is expected to have a negative impact on accuracy. Note that even in the ``No DP'' scenario, there is some privacy protection, as anonymity is assumed. The accuracy of this model proves the first hypothesis, namely that a predictor of vocal characteristics can be learned from only the voice trigger phrase.

  \item[Local DP] The strongest form of differential privacy, local DP, is applied using the Gaussian mechanism \cite{balle18}. The privacy parameters used were $\epsilon=2$ and $\delta=10^{-5}$. This results in a significant negative impact on the performance of the model, yielding an accuracy of $82.1\%$, with a low SNR of $0.19$ observed for the first central model update.
The result is still much better than random, which shows that useful knowledge can be learned even with such strict privacy guarantees and low SNR.
  \item[Central DP] The Gaussian moments accountant is applied on the aggregate model update to provide central privacy guarantees. The privacy parameters used were $\epsilon=2$ and $\delta=10^{-5}$. For the moments accountant, the population size is assumed to be $100$M, the cohort size is $300$ and the maximum number of central iterations is $60$. The accuracy of the trained vocal classification model is $94.6\%$, which is close to the ``No DP'' case. An SNR of $1.31$ is observed for the first central model update, which is significantly higher than ``Local DP''. However, this does not have any local privacy guarantees. 
  \item[Central DP with weaker local DP] Falling between the ``local DP'' and ``Central DP'' experiments, a weaker form of local DP \cite{bhowmick19} is used in combination with the Gaussian moments accountant for the central DP mechanism. This combination results in an accuracy of $94.1\%$, which is a good privacy-utility trade-off. The privacy parameters used in the weaker form of local DP translate to $\epsilon=25.7$, which is in the high-epsilon regime. However, with the assumed privacy amplification through shuffling and sampling for anonymity, and the application of central DP with $\epsilon=2$ and $\delta=10^{-5}$, this may be considered a reasonable operating point. The SNR observed here is $1.07$, which is significantly higher than ``local DP'' and expected considering the high $\epsilon$ value.
\end{description}

Hyperparameter tuning was performed in simulations with limited central data.
The best performing model was used as an initialization when training with real devices.
Switching to training the vocal classification model distributed on-device yielded multiple benefits.
Firstly, additional categories of side information not previously available in the limited central dataset were used. 
Secondly, magnitudes more data is available distributed on devices. The cohort of users for each central model update was increased to $5000$, resulting in higher SNR for the same local DP parameters and smaller noise variance from the moments accountant used centrally.
This larger effective corpus means that there is more speaker coverage.
Thirdly, while labels of the central dataset may be erroneous due to errors in manual human annotation, on-device training uses ground truth labels, thereby increasing accuracy, while protecting privacy.

\section{Multi-task learning of the speaker~verification~system}\label{sec:setup}

The third hypothesis proposed in this paper is that the encoded knowledge of a vocal classification model can act as complementary information for training a more accurate speaker verification system. 
Multiple approaches for utilizing the knowledge of the vocal classification model were experimented with: static rules for rejecting a request based on the model output, using the output of the model as input to intermediate layers when training the speaker embedding network, and multi-task learning with pseudo-labels.
The latter approach was the most successful and is the focus of the rest of the paper.

Specifically, the network was trained to predict the speaker, like the baseline system, and, additionally, the side information.
The loss was the sum of the original loss and a new term.
Minimizing this loss distilled the knowledge \cite{hinton15} of the vocal classification model, encoded in the pseudo-labels it generates.
Just as with the baseline system, the final classification layer was removed during inference to expose the embedding layer.
Not only did distilling the knowledge of the vocal classification model outperform the two other approaches mentioned, but the final architecture of the speaker verification system remains unchanged. 
Since the vocal classification model is a model trained with differential privacy, any knowledge that is distilled by the speaker embedding network is also protected by the post-processing theorem of differential privacy \cite{dwork14}.

The vocal classification model is generally confident in its predictions, mostly generating an output probability of $0.995$ for the highest predicted class, even where it is incorrect.
To better distill the knowledge for cases like this, the concept of temperature is used \cite{hinton15}. 
A temperature higher than $1$ softens the output distribution, making the probabilities that were previously minuscule more representative.

Given a mini-batch of data $\mathbf{X} = \{\mathbf{x}^{(t)}\}$ and corresponding labels $\mathbf{Y}= \{\mathbf{y}^{(t)}\}$, the objective function to minimize for the multi-task setup is

\vspace*{-0.2cm}
\begin{equation}
	\mathcal{L}_{mtl}(\mathbf{X},\mathbf{Y}) = \sum_{t} 
		\big(
			\mathcal{L}_{spk}(\mathbf{x}^{(t)}, \mathbf{y}^{(t)}) + 
			\gamma\mathcal{L}_{vc}(\mathbf{x}^{(t)}) 
		\big),
\end{equation}
where $\mathcal{L}_{spk}$ is the cross-entropy loss function for speaker identification as in the baseline setup, $\gamma$ is a weight for the vocal classification loss relative to speaker identification, and the vocal classification loss $\mathcal{L}_{vc}$ is defined as

\vspace*{-0.4cm}
\begin{equation}
	\mathcal{L}_{vc}(\mathbf{x}) = \frac{T^2}{N} \sum_{i=0}^N \Bigg( 
			\frac{e^{
				\mathcal{V}_i(\mathbf{x})/T
			}}{\sum_j e^{
				\mathcal{V}_j(\mathbf{x})/T			
			}}
			-
			\frac{e^{
				z_i/T
			}}{\sum_j e^{
				z_j/T			
			}}
	\Bigg)^2.
\end{equation}
Here, $z_i$ is the $i$th logit of the predicted side information from the vocal classification layer in the speaker embedding training setup, $\mathcal{V}_i(\mathbf{x})$ is the $i$th logit from forward propagating the supervector $\mathbf{x}$ through the vocal classification model and $T$ is the temperature. 
The above can be interpreted as the mean squared error between the ``softened'' softmax distributions. 
The mean squared error is multiplied by $T^2$ because otherwise $\frac{\partial \mathcal{L}_{vc}}{\partial z_i}$ scales with a factor of $\frac{1}{T^2}$ for large $T$ (see section 2.1 of \cite{hinton15}). 

\section{Results}\label{sec:results}

In this section, results of three models are presented, all with the same final network architecture. 
The difference is only in how the speaker embedding is trained. 
The first model, \textit{Baseline}, is the baseline production system described in Section \ref{sec:phs}. 
The second model, \textit{VC offline}, follows the setup described in Section \ref{sec:setup} where the knowledge to distil is from a vocal classification model trained on the limited offline data (blue line in Figure \ref{fig:vc}).
The third model, \textit{VC FL}, also follows the setup from Section \ref{sec:setup}, but with a vocal classification model trained with federated learning. The mechanism in \cite{bhowmick19} was used for local privacy guarantees and the Gaussian mechanism with moments accountant was used for central privacy guarantees.

Multi-task learning of the speaker embedding was conducted on $1.7$ million utterances of the trigger phrase from $18700$ speakers, all preprocessed by the voice trigger system to extract 520-dimensional supervectors. 
The softmax layer classifying side information has six outputs and the softmax layer classifying speakers has one output per speaker. 
The temperature $T$ was set to $10$ for all experiments, and the weight $\gamma$ of the side information classification loss was roughly tuned to balance the losses of the two tasks. 
Evaluation accuracy was measured on another set of $93600$ utterances from the same population of speakers as the training dataset.

The accuracies of speaker identification and side information classification on the evaluation dataset are shown in Table \ref{tab:phs-train}.
The accuracy on the speaker identification task in the multi-task setting increases relative to the baseline. 
It is possible that the side information has both a regularizing effect on speaker identification as well as helping propagate signals through the network.
The accuracy on the classification of the side information is expected to be close to $100\%$ because the labels are generated from the vocal classification model, and this larger DNN should be able to capture the encoded knowledge of the vocal classification model.

\begin{table}
  \caption{Performance of the two tasks in the multi-task learning setup on an evaluation set.}
  \label{tab:phs-train}
  \centering
  \begin{tabular}{ l  l  l  }
    \toprule
    \thead{Model} &  
    \thead{Speaker accuracy (\%)} &  
    \thead{Side information \\ accuracy (\%)} \\
    \midrule
    Baseline & 82.56\% & -~~~             \\
    VC offline & 83.37\% & 98.22\%~~~             \\
    VC FL & 83.59\% & 98.15\%~~~             \\
    \bottomrule
  \end{tabular}
\end{table}

As mentioned in Section \ref{sec:phs}, a speaker profile is defined by a set of supervectors from enrolment utterances. 
When evaluating the speaker verification system with the newly trained speaker embedding network, each of $234$ speaker profiles available were compared to supervectors of test utterances using Equation \ref{eq:predict}. 
A subset of the test utterances were unique utterances from the $234$ speakers which had profiles, and the rest originated from imposter speakers that did not match any profile. 
A total of $65545$ pairs of speaker profiles and test utterances were compared and a test utterance was accepted or rejected by applying a threshold $\theta$ on the speaker vector score $SV_{score}$. 
Performance at the equal error rate (EER) is shown in Table \ref{tab:phs-eval} for the three experiments. 
The multi-task setup with knowledge distillation of a vocal classification model trained on limited central data yields a $1.5\%$ absolute improvement over the baseline. The same setup with a vocal classification model trained on-device with privacy-preserving federated learning yields a $6\%$ relative improvement in EER. These results prove both our second and third hypotheses, namely that privacy-preserving federated learning can be used to improve the vocal classification-based system (due to the gain over the ``VC offline'' experiment), and that the vocal classification model can be used through multi-task learning to improve the speaker verification system (due to the gain over the ``Baseline'' system).
\begin{table}
  \caption{Performance of speaker verification on a test set.}
  \label{tab:phs-eval}
  \centering
  \begin{tabular}{ l  l }
    \toprule
    \thead{Model} &  
    \thead{EER (\%)}  \\
    \midrule
    Baseline & 10.10~~~             \\
    VC offline & 9.95~~~             \\
    VC FL & 9.50~~~             \\
    \bottomrule
  \end{tabular}
\end{table}

\section{Conclusions}

This paper demonstrates how a centrally trained speaker verification system can be improved by distilling the knowledge of an auxiliary model that was trained with side information on a much broader population using federated learning, while protecting user privacy.
Firstly, the auxiliary model, which classifies side information, was trained using data distributed over millions of real devices.
Additional experiments simulated different combinations of federated learning and differential privacy when training this model, to highlight the utility/privacy trade-off expected when using such approaches. The accuracy, time to convergence and signal-to-noise ratio clearly show the relative ordering of these approaches in terms of utility.
Secondly, the encoded knowledge of the auxiliary model was distilled into the speaker embedding network of an existing speaker verification baseline system using multi-task learning. 
Finally, a relative improvement of $6\%$ in equal error rate for speaker verification was achieved using this technique while maintaining the same network architecture as the baseline. This result shows that the speaker characteristic knowledge distilled into the speaker verification network resulted in speaker embeddings which are more discriminative.

\section{Acknowledgements}
This work was a collaborative effort between multiple teams. The authors would like to thank Chandra Dhir and Sachin Kajarekar for their help and involvement in relation to the speaker verification system. The authors would also like to thank everyone involved in the private federated learning effort, which made the experiments in this work possible including: Abhishek Bhowmick, Simon Beaumont, Andrew Byde, Luke Carlson, Andrew Cherkashyn, Mansi Deshpande, Fei Dong, Julien Freudiger, Stanley Hung, Omid Javidbakht, Gaurav Kapoor, Joris Kluivers, Henry Mason, Tom Naughton, Deepa Nemmili Veeravalli, Rehan Rishi and Dominic Telaar.

\bibliographystyle{IEEEtran}

\bibliography{references}

\end{document}